\documentclass[aps,prl,citeautoscript,showpacs,reprint,amsmath,amssymb]{revtex4-1}

\usepackage{graphicx}
\usepackage{mathtools}
\usepackage{dblfloatfix}
\usepackage{times}
\usepackage{bm}
\usepackage{ulem}
\usepackage[colorlinks,linkcolor=blue,citecolor=blue,pdfstartview=FitH,plainpages=false]{hyperref}
                        
\usepackage{color}
\usepackage{soul}

\newsavebox{\myfig}

\begin{document}

\title{From Ferromagnet to Antiferromagnet: Dimensional Crossover in (111) SrRuO$_3$ Ultrathin Films}

\author{Zhaoqing Ding$^{1,2}$}
\author{Xuejiao Chen$^{3}$}
\email{chenxj@czu.cn}
\author{Lei Liao$^{1,2}$}
\author{Zhen Wang$^{1,4}$}
\author{Zeguo Lin$^{1,2}$}
\author{Yuelong Xiong$^{1,2}$}
\author{Junzhou Wang$^{1,2}$}
\author{Fang Yang$^{1}$}
\author{Jiade Li$^{5}$}
\author{Peng Gao$^{5,6}$}
\author{Lifen Wang$^{1,2}$}
\author{Xuedong Bai$^{1,2,7}$}
\author{Xiaoran Liu$^{1}$}
\email{xiaoran.liu@iphy.ac.cn}
\author{Jiandong Guo$^{1,2}$}
\email{jdguo@iphy.ac.cn}

\affiliation{$^1$ Beijing National Laboratory for Condensed Matter Physics and Institute of Physics, Chinese Academy of Sciences, Beijing 100190, China.}
\affiliation{$^2$ School of Physical Sciences, University of Chinese Academy of Sciences, Beijing 100049, China.}
\affiliation{$^3$ School of Photoelectric Engineering, Changzhou Institute of Technology, Changzhou, Jiangsu 213002, China.}
\affiliation{$^4$ Institute of High Energy Physics, Chinese Academy of Sciences, Beijing 100049, China.}
\affiliation{$^5$ International Center for Quantum Materials, and Electron Microscopy Laboratory, School of Physics, Peking University, Beijing 100871, China.}
\affiliation{$^6$ Tsientang Institute for Advanced Study, Zhejiang 310024, China.}
\affiliation{$^7$ Songshan Lake Materials Laboratory, Dongguan, Guangdong 523808, China.}


\begin{abstract}
SrRuO$_3$ is a canonical itinerant ferromagnet, yet its properties in the extreme two-dimensional limit on a (111) crystal plane remain largely unexplored. Here, we demonstrate a complete transformation of its ground state driven by dimensional reduction. As the thickness of (111)-oriented SrRuO$_3$ films is reduced to a few unit cells, the system transitions from a metallic ferromagnet to a semiconducting antiferromagnet. This emergent antiferromagnetism is evidenced by a vanishing magnetic remanence and most strikingly, by the appearance of an unconventional twelve-fold anisotropic magnetoresistance. First-principles calculations confirm that an A-type antiferromagnetic order is the stable ground state in the ultrathin limit. Our findings establish (111) dimensional engineering as a powerful route to manipulate correlated electron states and uncover novel functionalities for antiferromagnetic spintronics.
\end{abstract}

\maketitle

Dimensionality effects remain a central and ongoing topic in condensed matter physics.\cite{1,2,3,4,5} Reduced dimensionality frequently induces profound modifications in structural and electronic properties of a system, as paradigmatically illustrated by the remarkable distinction between the three-dimensional (3D) graphite and its two-dimensional (2D) graphene. Recent advancements in the synthesis of high-quality epitaxial films have enabled precise control over growth condition, achieving atomically engineered quasi-2D artificial lattice motifs with the observation of exotic dimensionality-driven phenomena.\cite{6,7,8,9,10,11}
In this context, ultrathin films and heterostructures constituted by transition metal oxides (TMOs) have garnered widespread attention by exhibiting a wealth of novel physical properties, arising from the intricate coupling of spin, lattice, charge and orbital degrees of freedom. Prominent examples include two-dimensional electron gases (2DEGs),\cite{12,13} interface superconductivity,\cite{14} and emergent complex magnetism tailored by strain engineering. \cite{15,16,17,18,19}
In particular, theoretical investigations inspired from the honeycomb-lattice physics, propose that the perovskite structure can be naturally regarded as 3D extended honeycomb architectures along the [111] crystallographic axis. This intrinsic geometric organization, once confined to quasi-2D slabs, creates a unique platform for establishing a rich variety of correlated and topological phases of matter.\cite{20,21,22,23}
Notably, the (111)-terminated surface exhibits a hexagonal symmetry that is regarded as the key to realizing diverse nontrivial states,\cite{24} as schematized in Figure 1a.
Subsequent experiments have revealed that (111)-oriented perovskite oxide heterostructures host exotic quantum states fundamentally distinct from their bulk or (001)-oriented counterparts.\cite{25,26,27,28,29,30,31,32,33,34} 

SrRuO$_3$, driven by the itinerant and localized duality of Ru $4d$ electrons, exhibits an extraordinary itinerant ferromagnetism in the TMO compounds.\cite{35,36} 
While the general physical properties of SrRuO$_3$ have been extensively studied for long,\cite{35} this material has regained significant interest over the past decade for nontrivial topological features.   
For instance, its intrinsic Berry curvature was proposed through first-principles calculations about the non-monotonic relationship between anomalous Hall conductivity ($\sigma_{xy}$) and magnetization ($M$), the phenomenon later recognized as a hallmark signature of Weyl points in the electronic bands.\cite{37,38,39,40} Meanwhile, SrRuO$_3$ also serves as an exceptional playground for realizing topological spin textures in real space, such as skyrmions, via interfacial Dzyaloshinskii-Moriya interaction (DMI).\cite{41,42,43,44,45}
The influence of dimensionality effects in SrRuO$_3$ has predominantly focused on the (001) orientation, where confinement has been used to engineer magnetic anisotropy \cite{46}, manipulate the topological Hall effect,\cite{44,45,47} and stabilize high-spin-polarized 2DEGs.\cite{48} However, the [111] direction represents a largely uncharted frontier. Recent pioneering studies have unveiled a collection of intriguing phenomena, including the control of the magnetic Weyl fermion,\cite{34} the tensile-strain induced $R\bar{3}c$ symmetry and nonlinear transport,\cite{26} the emergence of a chiral spin crystal phase,\cite{31} and a thickness-driven metal-insulator transition in the ultrathin limit, though the latter finding is based preliminarily on transport probes.\cite{49} Nevertheless, a systematic investigation of how dimensionality modulates the electronic and magnetic properties in this archetypal ruthenate along the [111] direction remains fundamentally unexplored.


Here, we report on the dimensional reduction of SrRuO$_3$ along the [111] crystallographic direction. From 3D to quasi-2D limit, the (111)-oriented SrRuO$_3$ undergoes a metal-to-semiconductor transition, with its magnetic hysteresis loops showing characteristics reminiscent of antiferromagnetism. These SrRuO$_3$ ultrathin films exhibit an unconventional twelve-fold anisotropic magnetoresistance (AMR), in addition to the standard two- and six-fold harmonics. Stoner-Wohlfarth modeling attributes this higher-order periodicity to antiferromagnetic spin alignment, and first-principles calculations confirm that dimensional reduction along [111] stabilizes A-type antiferromagnetic (AFM) order as the ground state of SrRuO$_3$. Our findings highlight an opportunity to revealing elusive correlated-topology phenomena through precise (111) dimensional control.

\vskip 4mm

\begin{figure}[htp]
\centering
\includegraphics[width=0.49\textwidth]{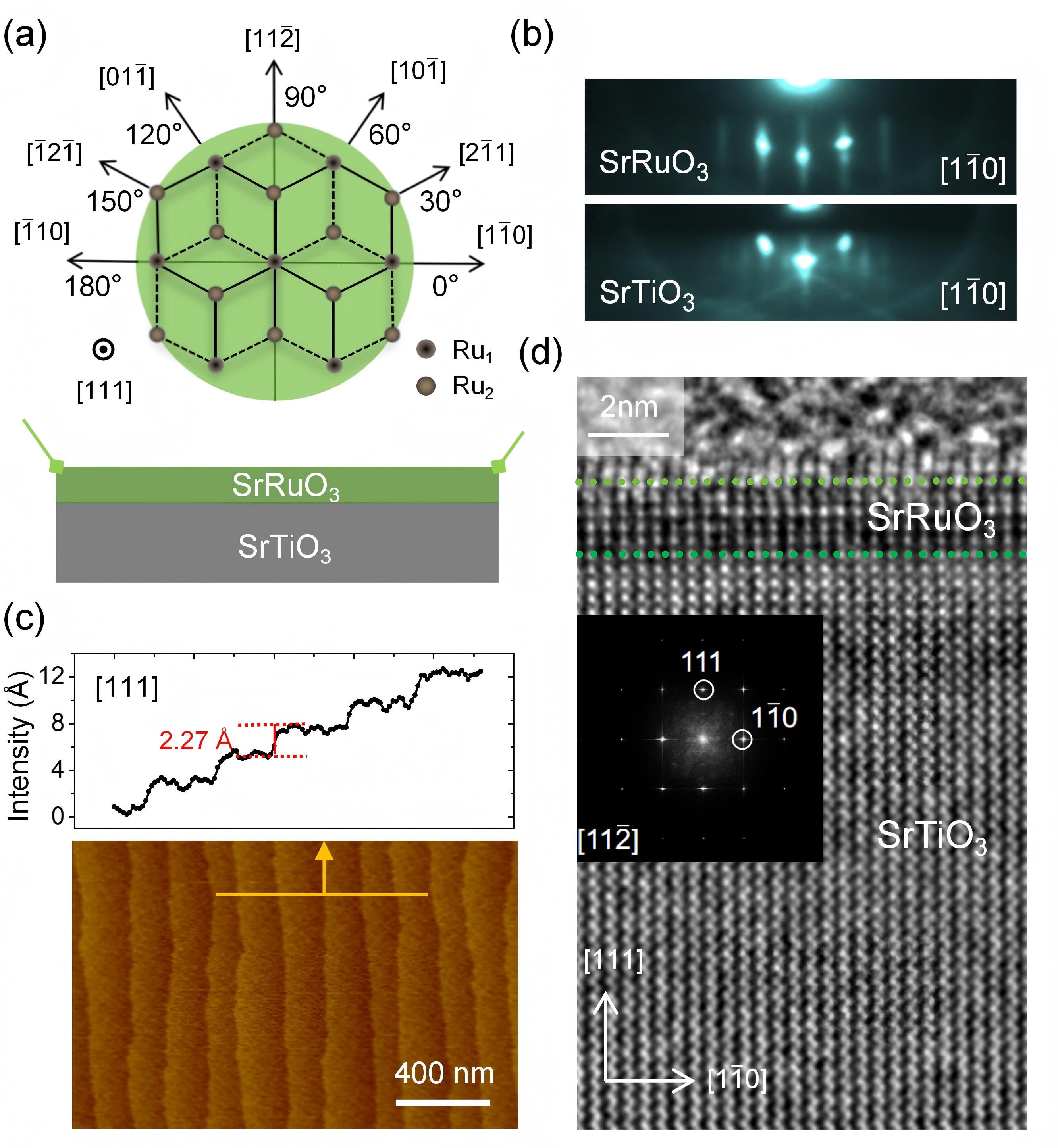}
\caption{\label{Fig1} Growth and structural characterization of (111) SrRuO$_3$ ultrathin films. (a) Special crystallographic directions within the (111) plane denoted in angle positions, with 0$^{\circ}$ defined as the [$\bar{1}$10] direction.
(b) {\it In-situ} RHEED patterns of SrTiO$_3$ substrate and 6 u.c. SrRuO$_3$ ultrathin films. The electron beam is incident along the [1$\bar{1}$0] direction.
(c) AFM graph on the surface morphology of 6 u.c. SrRuO$_3$ on Ti\textsuperscript{4+}-terminated SrTiO$_3$(111) substrate. The height of each step identified by line profile analysis is about 2.27 \AA.    
(d) Cross-sectional TEM image of the 6 u.c. SrRuO$_3$ projected along the [11$\bar{2}$] crystallographic direction. The inset shows the corresponding diffractogram of the SrTiO$_3$ substrate.}
\end{figure}

\medskip

A series of (111) SrRuO$_3$ thin films with different thicknesses ranging from 6 to 75 unit cell (u.c.) were fabricated using the pulsed laser deposition technique. 
We employed the (111)-oriented SrTiO$_3$ substrates, which provide exceptional lattice template to establish the 2D growth of SrRuO$_3$ (lattice mismatch $\approx$ 0.6\%).\cite{50,51}
The SrTiO$_3$ substrates were first thermally treated at $1150\,^{\circ}\mathrm{C}$ for 1.5 hours, giving rise to the formation of Ti\textsuperscript{4+}-terminated surface with well-defined terraces and steps. Prior to the deposition of SrRuO$_3$ thin films, a SrTiO$_3$ buffer layer of $\sim$2.2 nm was homoepitaxially grown on the SrTiO$_3$ substrate. Then SrRuO$_3$ was deposited at a substrate temperature of $800\,^{\circ}\mathrm{C}$ under an oxygen partial pressure of 10 Pa. Samples were annealed at the growth condition for 15 min and subsequently cooled to room temperature at a constant rate of 10\,\textdegree C\,/min.  
This sequential deposition process successfully produced epitaxial SrRuO$_3$ films of high crystallinity and flattness, as monitored via the {\it in-situ} reflection high-energy electron diffraction (RHEED) patterns presented in Figure 1b. 
Atomic force microscopy (AFM) imaging shown in Figure 1c confirmed the surface morphology of SrRuO$_3$ films with well-established terraces and steps. The height of each step is about 2.27 \AA, in accord with the value of (111) lattice spacing of 1 u.c. SrRuO$_3$.\cite{52}
Cross-sectional transmission electron microscopy (TEM) image shown in Figure 1d on a 6 u.c. SrRuO$_3$ further verified the expected epitaxial relationship between SrRuO$_3$ film and SrTiO$_3$ substrate with abrupt interface.



\begin{figure}[htp]
\centering
\includegraphics[width=0.48\textwidth]{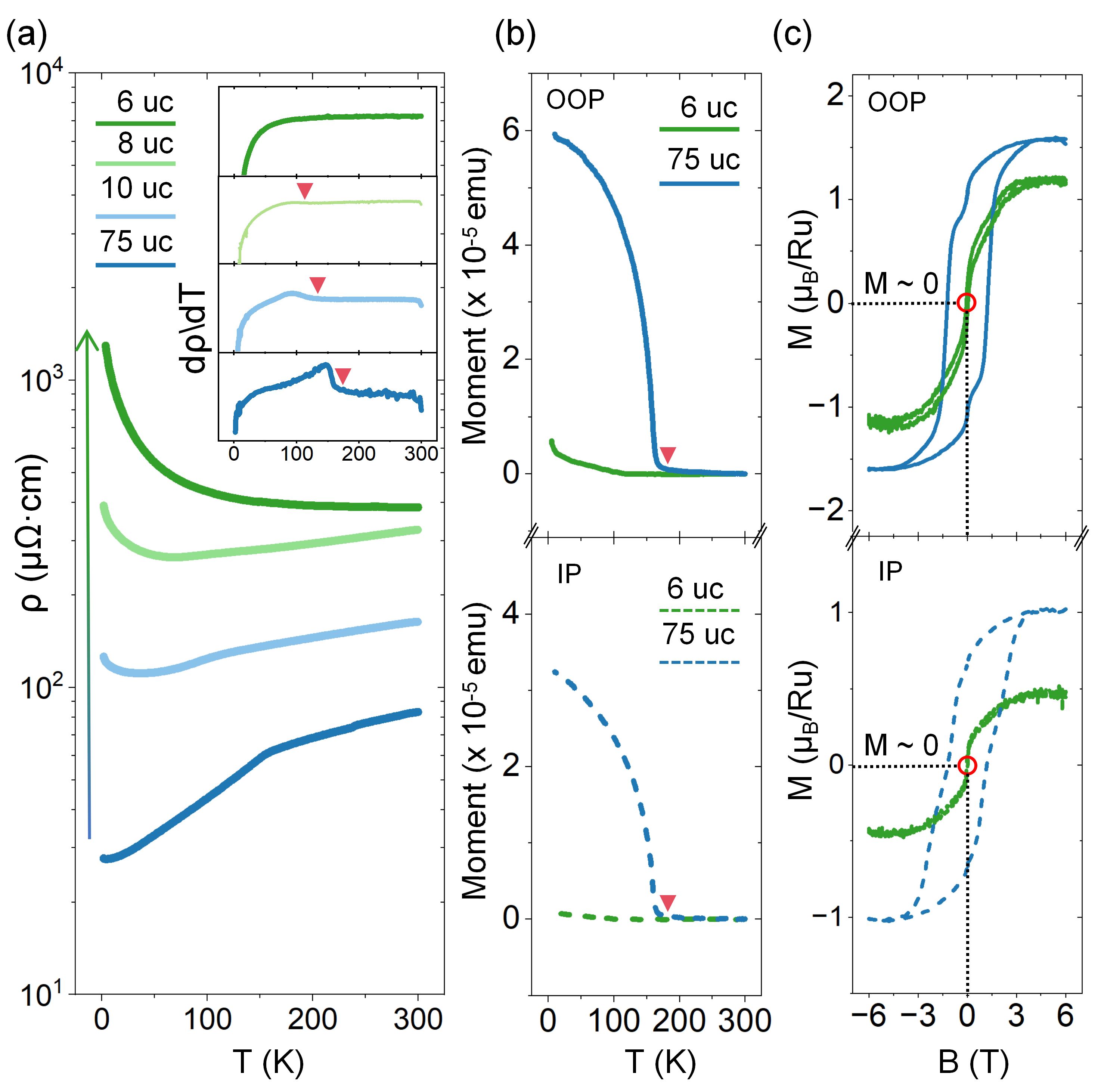}
\caption{\label{Fig2} Electrical and magnetic properties of (111) SrRuO$_3$ films. (a) Temperature-dependent resistivity curves of SrRuO$_3$ films (6 - 75 u.c.). Inset: First derivative $d\rho / dT$ with the kink temperatures marked by red triangles.
(b) Temperature-dependent magnetization curves of both 6 and 75 u.c. SrRuO$_3$ after field cooling (FC) under 2000~Oe magnetic field for both IP and OOP orientations.
(c) Isothermal magnetization hysteresis loops at 4~K for IP and OOP field orientations.}
\end{figure}
\medskip

After demonstrating the high quality of films, we focus on their electrical transport and magnetic properties. 
As shown in Figure 2a, the temperature-dependent resistivities of our films indicate a metal-semiconductor (or metal-semimetal) crossover when decreasing SrRuO$_3$ film thickness down to 6 u.c. ($\sim$1.36 nm), qualitatively similar to previous results.\cite{49,53,54,55} Notably, this thickness is only half of that reported by Rastogi {\it et al.}, where a metal-insulator transition occurs below a critical thickness of $\sim$2.7 nm in (111) SrRuO$_3$ ultrathin films.\cite{49} The suppressed thickness threshold and enhanced conductivity of our ultrathin films likely benefits from the precise control of growth, which mitigates extrinsic issues such as defects or interfacial mixing. \cite{9,10,11}
The inset of Figure 2a displays the differential resistivity $d\rho / dT$ as a function of temperature for our SrRuO$_3$ films. The para- to ferro-magnetic (FM) transition, manifested as a kink on $d\rho / dT$ (marked by the red triangles), decreases monotonically with film thinning and vanishes in 6 u.c. SrRuO$_3$, while the 75 u.c. SrRuO$_3$ maintains \(T_{\mathrm{C}}\) near 155 K, consistent with its bulk value.\cite{35,36}

More knowledge about the magnetic nature of our (111) SrRuO$_3$ ultrathin films are obtained through magnetization measurements. As shown in Figure 2b, both the in-plane (IP) and out-of-plane (OOP) temperature-dependent magnetization $M(T)$ curves of 75 u.c. SrRuO$_3$ reveal a Curie temperature at 155 K, in good agreement with the results from resistivity measurements. However, for 6 u.c. SrRuO$_3$ exhibiting the semiconducting (or semimetal) behavior, no characteristic of FM transition is practically observed. 
Such a difference may suggest that the dimensional confinement induces a novel magnetic ground state in (111) SrRuO$_3$ ultrathin films beyond the conventional FM ordering. 
The field-dependent magnetization $M(H)$ curves provide additional evidences for this speculation. As seen in Figure 2c, 
For 75 u.c. SrRuO$_3$, clear hysteresis behaviors are detected on both IP and OOP $M(H)$ curves. Specifically, the OOP loops exhibit a saturated magnetization $M_s$ around 1.6 $\mu_{\mathrm{B}}$/Ru and a zero-field remanence $M_0$ around 1.2 $\mu_{\mathrm{B}}$/Ru, while the IP counterparts display reduced values with $M_s$ around 1.2 $\mu_{\mathrm{B}}$/Ru and $M_0$ around 0.8 $\mu_{\mathrm{B}}$/Ru. 
This OOP-dominated magnetic response reflects preferential orientation of the magnetic easy-axis toward the OOP direction, consistent with previous reports. \cite{56} 
However, in stark contrast, 6 u.c. SrRuO$_3$ exhibits remarkably distinct features: 1. Both the IP and OOP hysteretic loops are much narrower with reduced values of $M_s$ compared to 75 u.c. SrRuO$_3$; 2. More essentially, the complete absence of finite remanent magnetization $M_0$, as highlighted by the open red circles at zero field, shed light on its AFM nature. 


\begin{figure}[htp]
\centering
\includegraphics[width=0.49\textwidth]{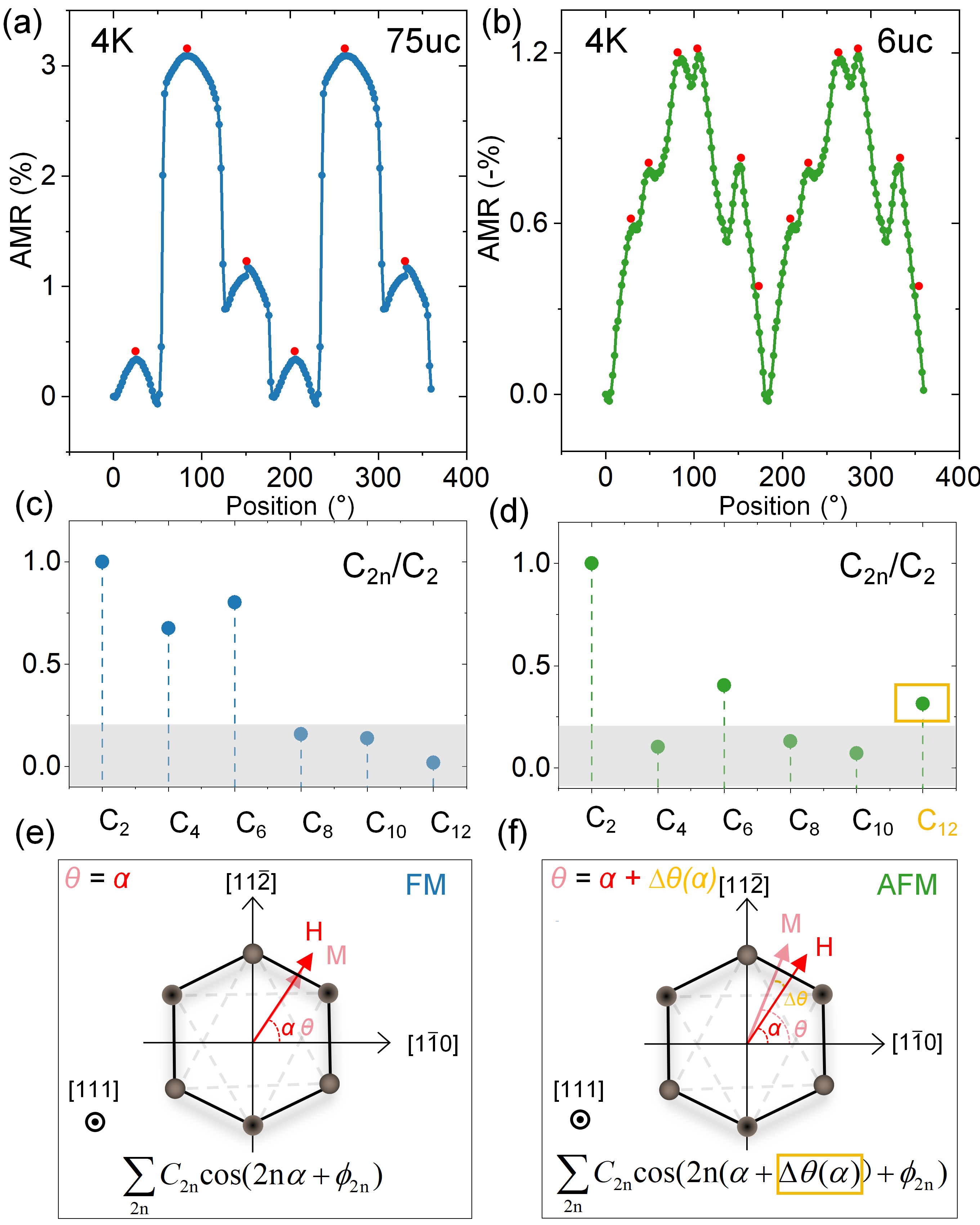}
\caption{\label{Fig3} In-plane AMR of both 75 and 6 u.c. SrRuO$_3$ with Fourier analysis and Stoner-Wohlfarth modelling. (a) AMR data of the 75 u.c. SrRuO$_3$ at 4 K and 5 T magnetic field. The peak positions are indicated in red.
(b) AMR data of the 6 u.c. SrRuO$_3$ at 4 K and 5 T magnetic field. The peak positions are indicated in red. 
(c) Amplitude of each harmonics $C_{2n}$ normalized to $C_2$ extracted from Fourier analysis.
(d) Amplitude of each harmonics $C_{2n}$ normalized to $C_2$ extracted from Fourier analysis. The enhanced magnitude of the twelve-fold $C_{12}$ is highlighted in orange. 
(e) Stoner Wohlfarth modeling on buckled honeycomb lattice for collinear $\mathbf{H}\parallel\mathbf{M}$ configuration.
(f) Stoner Wohlfarth modeling on buckled honeycomb lattice for nonparallel $\mathbf{H}$--$\mathbf{M}$ configuration with a small misalignment angle $\Delta\theta\neq 0$.}
\end{figure}

\medskip

Next, we turn to AMR to probe the spin configuration of the emergent magnetic state in (111) ultrathin SrRuO$_3$. AMR originates from spin-orbit coupling (SOC) and is exquisitely sensitive to the interplay between crystalline and magnetic symmetries \cite{57,58,59}. While conventional ferromagnets typically show two-fold $C_2$, four-fold $C_4$, and six-fold $C_6$ AMR harmonics reflecting the underlying crystal structure \cite{60,61,62}, recent studies have identified higher-order harmonics (e.g., eight- or twelve-fold) as definitive fingerprints of antiferromagnetism, where the interplay between the crystal lattice and the magnetic sublattice orientation folds the rotational symmetry of the N\'{e}el vector, halving the fundamental period of the AMR response \cite{63,64,65}.

Our measurements reveal a stark contrast between the thick and ultrathin films. The AMR of the 75 u.c. film exhibits the expected six-fold symmetry, with peaks appearing every $60^{\circ}$ (Figure~3a). In the 6 u.c. film, however, the AMR profile becomes significantly more complex, characterized by the emergence of additional peaks that indicate a fundamental change in symmetry (Figure~3b).
To quantify this change, we performed a fast Fourier-transform (FFT) decomposition of the AMR data, with the resulting harmonic amplitudes normalized to the leading two-fold term (i.e. $C_{2n}/C_2$) shown in Figures~3c and 3d \cite{26,57,58}. For the 75 u.c. film, the AMR is, as expected, dominated by the $C_2$, $C_4$, and $C_6$ components, reflecting conventional AMR, the bulk pseudo-cubic lattice symmetry, and the (111) plane's six-fold anisotropy, respectively~\cite{62, 66, 67}. In sharp contrast, the 6 u.c. film shows a profound suppression of the four-fold $C_4$ contribution and the remarkable emergence of a twelve-fold $C_{12}$ harmonic, which becomes comparable in magnitude to the $C_6$ term.

\begin{table}[ht]
\centering
\caption{Extracted parameters from FFT analysis for 75 and 6 u.c. SrRuO$_3$ at 4 K.}
\label{tab:fft_parameters}
\begin{tabular}{|c|c|c|c|}
\hline
Parameter & 75 u.c. ($\Omega$) & Parameter & 6 u.c. ($\Omega$) \\ \hline
$C_{0}$ & 13.786 & $C_{0}$ & 82867 \\ \hline
$C_{2}$ & 0.151 & $C_{2}$ & 321 \\ \hline
$C_{4}$ & 0.102 & $C_{4}$ & 33 \\ \hline
$C_{6}$ & 0.121 & $C_{6}$ & 130 \\ \hline
$C_{8}$ & 0.024 & $C_{8}$ & 42 \\ \hline
$C_{10}$ & 0.021 & $C_{10}$ & 23 \\ \hline
$C_{12}$ & 0.006 & $C_{12}$ & 101 \\ \hline
\end{tabular}
\end{table}

\medskip

The appearance of a prominent $C_{12}$ harmonic is a powerful indicator of an AFM state. To understand this, the orientation of the N\'{e}el vector is critical. The very existence of a large, rotating AMR signal is compelling evidence that the N\'{e}el vector lies predominantly within the (111) plane, as an out-of-plane N\'{e}el vector would remain static against a rotating in-plane field and thus could not produce angular-dependent AMR.
Accordingly, we employ the Stoner-Wohlfarth-type modeling adapted for a two-sublattice, in-plane antiferromagnet governed by the energy density  \cite{68,69,70,71}:
\begin{equation}
\frac{E}{V}=J_{1}\mathbf{M_{1}}\cdot \mathbf{M_{2}}+E_{MAE}(\psi_{1})+E_{MAE}(\psi_{2})-\mu_{0}\mathbf{H}\cdot(\mathbf{M_{1}}+\mathbf{M_{2}})
\label{eq:energy}
\end{equation}
Here, $J_1$ is the AFM exchange constant and $E_{MAE} = K_{MAE}\sin^{2}(3\psi)$ is the six-fold magnetocrystalline anisotropy energy (MAE). As detailed in the Supplementary Material (SM), an in-plane field $\mathbf{H}$ (at angle $\alpha$) cants the sublattices, but the N\'{e}el vector (at angle $\theta$) does not rigidly follow the field. Instead, it develops a small angular deviation $\Delta\theta(\alpha) = \theta - \alpha$. The leading anisotropic term of AMR depends on the N\'{e}el vector's orientation as $C_6\cos(6\theta)$. Substituting for $\theta$ yields an expression dependent on the field angle $\alpha$, $\Delta R(\alpha) \approx C_6\cos[6(\alpha+\Delta\theta(\alpha))]$.
A second-order Taylor expansion of this expression, shown in the SM, naturally generates a twelve-fold harmonic, given by the nonlinear relationship between $\Delta \theta(\alpha)$ and $\alpha$: 
\begin{equation}
\Delta\theta(\alpha)\approx -\frac{\mu_{0}^{2}H^{2}}{12J_{1}K_{MAE}M}\frac{\sin\alpha\cos\alpha}{\cos6\alpha}
\end{equation}
The model also yields the amplitude of this emergent harmonic directly tied to the six-fold term:
\begin{equation}
C_{12}\approx C_{6}\left(\frac{1}{4}\lambda^{2}\right) \quad \text{where} \quad \lambda=\frac{\mu_{0}^{2}H^{2}}{12J_{1}K_{MAE}M}
\label{eq:c12}
\end{equation}

This result provides a quantitatively coherent explanation for our observations, cementing the conclusion that the emergent $C_{12}$ term is a direct consequence of the AFM N\'{e}el vector's nonlinear response to the external field within the six-fold crystal potential, as schematically illustrated in Figures 3e and 3f. The synergistic integration of AMR analysis with three definitive magnetic characteristics significant hysteresis loop contraction, pronounced suppression of saturation magnetization, and critical vanishing of remnant magnetization provides a convergent experimental confirmation that dimensional confinement in (111)-oriented SrRuO$_3$ drives a transition from ferromagnetic order to an emergent in-plane antiferromagnetic configuration.

\begin{figure}[htp]
\centering
\includegraphics[width=0.48\textwidth]{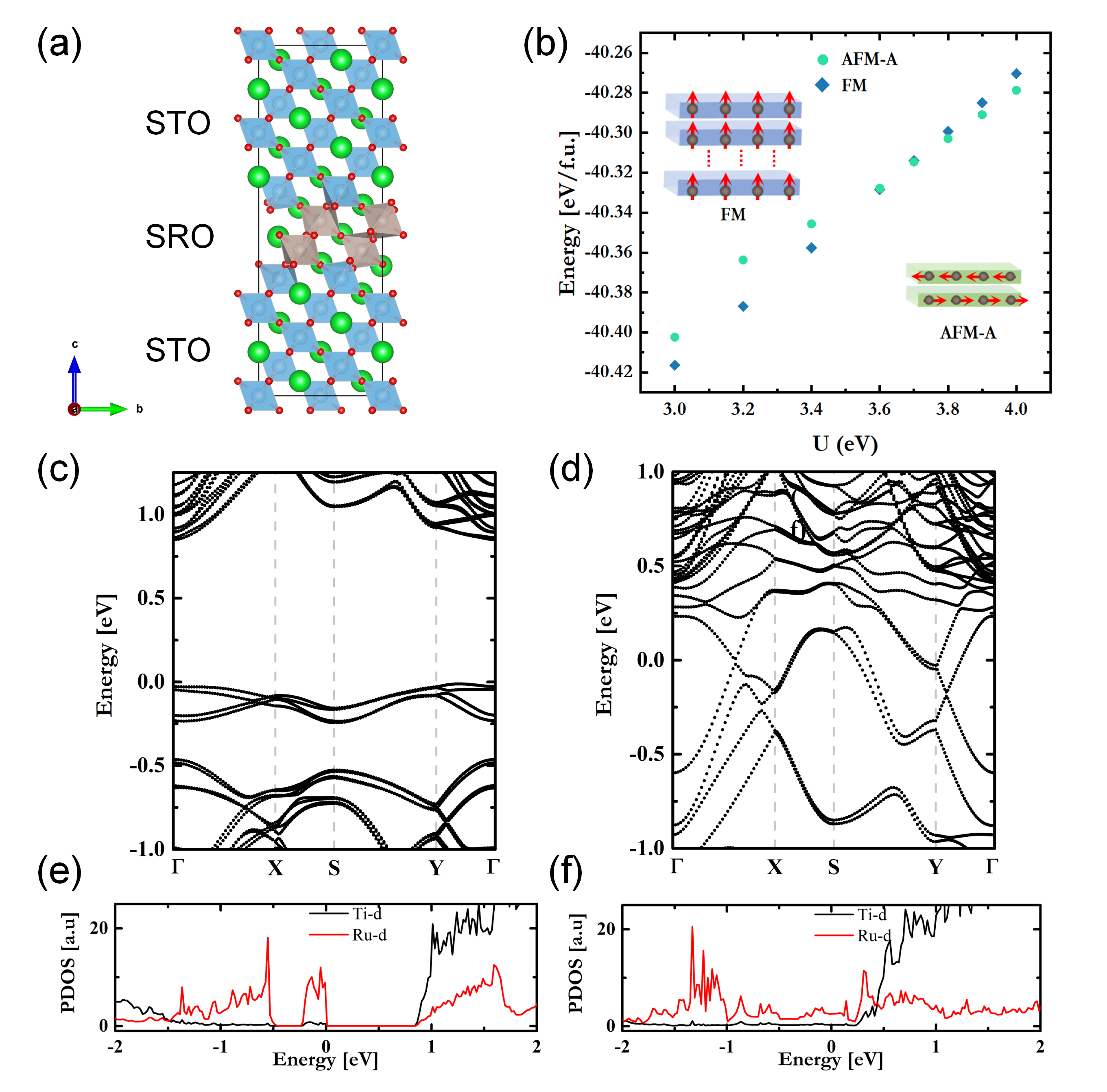}
\caption{\label{Fig4} The structural and electronic properties of  SrTiO$_3$-SrRuO$_3$ (111) directional superlattice based on GGA+U method. (a) The (111) directional superlattice of STO-SRO with layer ratio equal to 5:1.
(b) The calculated energies of AFM-A type and FM magnetic configuration along with U value from 3 to 4 eV. The turning point for AFM-A as the ground state appears around 3.7 eV.
The band structures of (c) AFM-A type with spin along the IP and (d) FM with spin along c axis. With the varied magnetic state, the conductivity changes from the metallic state of FM to the insulating result of AFM. The corresponding projected density of states for (e) AFM-A type and (f) FM configuration. }
\end{figure}

To provide theoretical corroboration for our experimental findings, we performed first-principles density functional theory (DFT) calculations \cite{72,73,74,75} on a (111)-oriented superlattice composed of 2 u.c. SrRuO$_3$ and 10 u.c. SrTiO$_3$ (Figure~4a). We first investigated the magnetic ground state stability, which in correlated $d$-electron systems is sensitive to the on-site Hubbard interaction $U$. As shown in Figure 4b, our calculations reveal a magnetic phase transition as a function of $U$: for $U < 3.7$~eV, the FM state is energetically favored, while for \textbf{$U > 3.7$~eV}, the A-type AFM state (AFM-A) becomes the stable ground state. From an experimental perspective, a large $U$ with a reduced bandwidth ($W$) strongly indicates that carrier localization in 6 u.c. SrRuO$_3$ from dimensional confinement effects.  This critical value is well within the range of accepted $U$ values for ruthenates, confirming the theoretical plausibility of an emergent AFM phase. Crucially, further calculations including SOC (see SM Table~I) confirm that an in-plane AFM-A configuration is the true ground state, in perfect agreement with the conclusions from our AMR analysis.

With the in-plane AFM-A order established as the theoretical ground state, we examined its electronic properties. The stable AFM-A configuration (Figure 4c) opens a distinct energy gap at the Fermi level. In striking contrast, the calculated band structure for the FM state (Figure 4d) exhibits highly dispersive Ru bands crossing the Fermi level, consistent with a metallic character. This calculated gap provides a direct theoretical basis for the semiconducting behavior observed in the transport measurements of our 6 u.c. film. The projected density of states displayed in Figures 4e and 4f confirms that the states near the Fermi level in both configurations are dominated by Ru-4$d$ orbitals. 

In summary, the combination of experiment vanishing remanent magnetization, emergent twelve-fold AMR and DFT calculations demonstrates that (111)-oriented dimensional confinement drives itinerant SrRuO$_3$ from the bulk itinerant ferromaget to a novel semiconducting antiferromagnet in ultrathin films. These findings establish universal design principles for low dimensional manipulation of perovskite oxide heterostructures via (111) heteroepitaxial engineering, and open new pathways for developing AFM spintronic devices operating in the quasi-2D quantum confinement regime. 

\medskip
{\it Acknowledgements.} 
This work is supported by the National Natural Science Foundation of China (No. 12204521, 12250710675, 12504198), and the National Key R\&D Program of China (No. 2022YFA1403000). A portion of this work was carried out at the Synergetic Extreme Condition User Facility (SECUF). A portion of this work was based on the data obtained at beamline 1W1A of Beijing Synchrotron Radiation Facility (BSRF-1W1A). A portion of this work was carried out at Electron Microscopy Laboratory of Peking University for the use of electron microscopes.\\


\begin{thebibliography}{99}\footnotesize
\itemsep=-1pt plus.2pt minus.2pt
\bibitem {1} Tsui D C, Stormer H L and Gossard A C 1982 {\it Phys. Rev. Lett.} {\bf 48} 1559
\bibitem {2} Nayak C 2010 {\it Nature} {\bf 464} 693
\bibitem {3} Mannhart J and Schlom D G 2010 {\it Science} {\bf 327} 1607
\bibitem {4} Boris A V, Matiks Y, Benckiser E, Frano A, Popovich P, Hinkov V, Wochner P, Castro-Colin M, Detemple E, Malik V K, Bernhard C, Prokscha T, Suter A, Salman Z, Morenzoni E, Cristiani G, Habermeier H-U and Keimer B 2011 {\it Science} {\bf 332} 937
\bibitem {5} Li L, Kim J, Jin C, Ye G J, Qiu D Y, da Jornada F H, Shi Z, Chen L, Zhang Z, Yang F, Watanabe K, Taniguchi T, Ren W, Louie S G, Chen X H, Zhang Y and Wang F 2017 {\it Nat. Nanotechnol.} {\bf 12} 21

\bibitem {6} Sun J Z, Abraham D W, Rao R A and Eom C B 1999 {\it Appl. Phys. Lett.} {\bf 74} 3017
\bibitem {7} Scherwitzl R, Gariglio S, Gabay M, Zubko P, Gibert M and Triscone J-M 2011 {\it Phys. Rev. Lett.} {\bf 106} 246403
\bibitem {8} Boschker I, Verbeeck J, Egoavil R, Bals S, van Tendeloo G, Huijben M, Houwman E P, Koster G, Blank D H A and Rijnders G 2012 {\it Adv. Funct. Mater.} {\bf 22} 2235
\bibitem {9} Ali Z, Wang Z, O'Hara A, Saghayezhian M, Shin D, Zhu Y, Pantelides S T and Zhang J 2022 {\it Phys. Rev. B} {\bf 105} 054429
\bibitem {10} Zhang X, Penn A N, Wysocki L, Zhang Z, van Loosdrecht P H M, Kornblum L, LeBeau J M, Lindfors-Vrejoiu I and Kumah D P 2022 {\it APL Mater.} {\bf 10} 051107
\bibitem {11} Wang G, Wang Z, Meng M, Saghayezhian M, Chen L, Chen C, Guo H, Zhu Y, Plummer E W and Zhang J 2019 {\it Phys. Rev. B} {\bf 100} 155114

\bibitem {12} Ohtomo A and Hwang H 2004 {\it Nature} {\bf 427} 423
\bibitem {13} Zhang H, Yun Y, Zhang X, Zhang H, Ma Y, Yan X, Wang F, Li G, Li R, Khan T, Chen Y, Liu W, Hu F, Liu B, Shen B, Han W and Sun J 2018 {\it Phys. Rev. Lett.} {\bf 121} 116803

\bibitem {14} Reyren N, Thiel S, Caviglia A D, Kourkoutis L F, Hammerl G, Richter C, Schneider C W, Kopp T, Ruetschi A S, Jaccard D, Gabay M, Muller D A, Triscone J M and Mannhart J 2007 {\it Science} {\bf 317} 1196

\bibitem {15} Huang B, Clark G, Navarro-Moratalla E, Klein D R, Cheng R, Seyler K L, Zhong D, Schmidgall E, McGuire M A, Cobden D H, Yao W, Xiao D, Jarillo-Herrero P and Xu X 2017 {\it Nature} {\bf 546} 270
\bibitem {16} Klein D R, MacNeill D, Lado J L, Soriano D, Navarro-Moratalla E, Watanabe K, Taniguchi T, Manni S, Canfield P, Fernandez-Rossier J and Jarillo-Herrero P 2018 {\it Science} {\bf 360} 1218
\bibitem {17} K. Ueda, H. Tabata, and T. Kawai, 1998 {\it Science} {\bf 280} 1064.
\bibitem {18} L. Li, C. Richter, J. Mannhart, and R. C. Ashoori, 2011 {\it Nat. Phys.} {\bf 7} 762.
\bibitem {19} A. J. Grutter, H. Yang, B. J. Kirby, M. R. Fitzsimmons, J. A. Aguiar, N. D. Browning, C. A. Jenkins, E. Arenholz, V. V. Mehta, U. S. Alaan, and Y. Suzuki, 2013 {\it Phys. Rev. Lett.} {\bf 111} 087202.

\bibitem {20}Haldane F D M 1988 {\it Phys. Rev. Lett.} {\bf 61} 2015
\bibitem {21}Xiao D, Zhu W, Ran Y, Nagaosa N and Okamoto S 2011 {\it Nat. Commun.} {\bf 2} 596
\bibitem {22}Wang F and Ran Y 2011 {\it Phys. Rev. B} {\bf 84} 241103
\bibitem {23}R\"{u}egg A and Fiete G A 2011 {\it Phys. Rev. B} {\bf 84} 201103

\bibitem {24}Rout P K, Agireen I, Maniv E, Goldstein M and Dagan Y 2017 {\it Phys. Rev. B} {\bf 95} 241107

\bibitem {25}Rastogi A, Brahlek M, Ok J M, Liao Z, Sohn C, Feldman S and Lee H N 2019 {\it APL Mater.} {\bf 7} 091106
\bibitem {26}Ding Z, Chen X, Wang Z, Zhang Q, Yang F, Bi J, Lin T, Wang Z, Wu X, Gu M, Meng M, Cao Y, Gu L, Zhang J, Zhong Z, Liu X and Guo J 2023 {\it npj Quantum Mater.} {\bf 8} 43
\bibitem {27}Gray B, Lee H N, Liu J, Chakhalian J and Freeland J W 2010 {\it Appl. Phys. Lett.} {\bf 97} 013105
\bibitem {28}Gibert M, Zubko P, Scherwitzl R, Iniguez J and Triscone J-M 2012 {\it Nat. Mater.} {\bf 11} 195
\bibitem {29}Liu X, Choudhury D, Cao Y, Middey S, Kareev M, Meyers D, Kim J-W, Ryan P and Chakhalian J 2015 {\it Appl. Phys. Lett.} {\bf 106} 071603
\bibitem {30}Hirai D, Matsuno J and Takagi H 2015 {\it APL Mater.} {\bf 3} 041508
\bibitem {31}Ding Z, Xie Y, Wang S, Chen X, Wang Z, Lin Z, Wang E, Wu X, Yang M, Xiong Y, Meng M, Yang F, Zhang J, Qiu X, Liu X and Guo J {\it unpublished}
\bibitem {32}Middey S, Meyers D, Doennig D, Kareev M, Liu X, Cao Y, Yang Z, Shi J, Gu L, Ryan P J, Pentcheva R, Freeland J W and Chakhalian J 2016 {\it Phys. Rev. Lett.} {\bf 116} 056801
\bibitem {33}Liu X, Fang S, Fu Y, Ge W, Kareev M, Kim J, Choi Y, Karapetrova E, Zhang Q, Gu L, Choi E, Wen F, Wilson J, Fabbris G, Ryan P, Freeland J, Haskel D, Wu W, Pixley J and Chakhalian J 2021 {\it Phys. Rev. Lett.} {\bf 127} 277204
\bibitem {34}Lin W, Liu L, Liu Q, Li L, Shu X, Li C, Xie Q, Jiang P, Zheng X, Guo R, Lim Z, Zeng S, Zhou G, Wang H, Zhou J, Yang P, Ariando, Pennycook S J, Xu X, Zhong Z, Wang Z and Chen J 2021 {\it Adv. Mater.} {\bf 33} 2101316

\bibitem {35} Koster G, Klein L, Siemons W, Rijnders G, Dodge J S, Eom C B, Blank D H A and Beasley M R 2012 {\it Rev. Mod. Phys.} {\bf 84} 253
\bibitem {36} Hahn S, Sohn B, Kim M, Kim J R, Huh S, Kim Y, Kyung W, Kim M, Kim D, Kim Y, Noh T W, Shim J H and Kim C 2021 {\it Phys. Rev. Lett.} {\bf 127} 256401

\bibitem {37} Berry M 1984 {\it Proc. R. Soc. London Ser. A} {\bf 392} 45
\bibitem {38} Fang Z, Nagaosa N, Takahashi K S, Asamitsu A, Mathieu R, Ogasawara T, Yamada H, Kawasaki M, Tokura Y and Terakura K 2003 {\it Science} {\bf 302} 92
\bibitem {39} Itoh S, Endoh Y, Yokoo T, Ibuka S, Park J-G, Kaneko Y, Takahashi K S, Tokura Y and Nagaosa N 2016 {\it Nat. Commun.} {\bf 7} 11788
\bibitem {40} Takiguchi K, Wakabayashi Y K, Irie H, Krockenberger Y, Otsuka T, Sawada H, Nikolaev S A, Das H, Tanaka M, Taniyasu Y and Yamamoto H 2020 {\it Nat. Commun.} {\bf 11} 4969

\bibitem {41} Wang L, Feng Q, Kim Y, Kim R, Lee K H, Pollard S D, Shin Y J, Zhou H, Peng W, Lee D, Meng W, Yang H, Han J H, Kim M, Lu Q and Noh T W 2018 {\it Nat. Mater.} {\bf 17} 1087
\bibitem {42} Meng K-Y, Ahmed A S, Ba\'{c}ani M, Mandru A-O, Zhao X, Bagu\'{e}s N, Esser B D, Flores J, McComb D W, Hug H J and Yang F 2019 {\it Nano Lett.} {\bf 19} 3169
\bibitem {43} Matsuno J, Ogawa N, Yasuda K, Kagawa F, Koshibae W, Nagaosa N, Tokura Y and Kawasaki M 2016 {\it Sci. Adv.} {\bf 2} e1600304
\bibitem {44} Wang H, Dai Y, Liu Z, Xie Q, Liu C, Lin W, Liu L, Yang P, Wang J and Venkatesan T V 2020 {\it Adv. Mater.} {\bf 32} 1904415
\bibitem {45} Seddon S D, Dogaru D E, Holt S J R, Rusu D, Peters J J P, Sanchez A M and Alexe M 2021 {\it Nat. Commun.} {\bf 12} 2007

\bibitem {46} Jeong S G, Cho S W, Song S, Oh J Y, Jeong D G, Han G, Jeong H Y, Mohamed A Y, Noh W-s, Park S, Lee J S, Lee S, Kim Y-M, Cho D-Y and Choi W S 2024 {\it Nano Lett.} {\bf 24} 7979

\bibitem {47} Cho S W, Jeong S G, Kwon H Y, Song S, Han S, Han J H, Park S, Choi W S, Lee S and Choi J W 2021 {\it Acta Mater.} {\bf 216} 117153

\bibitem {48} Verissimo-Alves M, Garc\'{l}a-Fern\'{a}ndez P, Bilc D I, Ghosez P and Junquera J 2012 {\it Phys. Rev. Lett.} {\bf 108} 107003

\bibitem {49} Rastogi A, Brahlek M, Ok J M, Liao Z, Sohn C, Feldman S and Lee H N 2019 {\it APL Mater.} {\bf 7} 091106

\bibitem {50} Lee J J, Schmitt F T, Moore R G, Johnston S, Cui Y-T, Li W, Yi M, Liu Z K, Hashimoto M, Zhang Y, Lu D H, Devereaux T P, Lee D-H and Shen Z-X 2014 {\it Nature} {\bf 515} 245
\bibitem {51} Wang J, Neaton J B, Zheng H, Nagarajan V, Ogale S B, Liu B, Viehland D, Vaithyanathan V, Schlom D G, Waghmare U V, Spaldin N A, Rabe K M, Wuttig M and Ramesh R 2003 {\it Science} {\bf 299} 1719

\bibitem {52} Lee J, Choi J K, Moon S Y, Park J, Kim J-S, Hwang C S, Baek S-H, Choi J-H and Chang H J 2015 {\it Appl. Phys. Lett.} {\bf 106} 071601


\bibitem {53} Grutter A J, Wong F J, Arenholz E, Vailionis A and Suzuki Y 2012 {\it Phys. Rev. B} {\bf 85} 134429
\bibitem {54} Lee C, Kwon O U, Shin R H, Jo W and Jung C U 2014 {\it Nanoscale Res. Lett.} {\bf 9} 8
\bibitem {55} Rastogi A, Brahlek M, Ok J M, Liao Z, Sohn C, Feldman S and Lee H N 2019 {\it APL Mater.} {\bf 7} 091106

\bibitem {56} Wang H, Laskin G, He W, Boschker H, Yi M, Mannhart J and van Aken P A 2022 {\it Adv. Funct. Mater.} {\bf 32} 2108475

\bibitem {57} McGuire T and Potter R 1975 {\it IEEE Trans. Magn.} {\bf 11} 1018
\bibitem {58} Ritzinger P and V\'{y}born\'{y} K 2023 {\it R. Soc. Open Sci.} {\bf 10} 230564

\bibitem {59} D\"{o}ring W 1938 {\it Ann. Phys.} {\bf 5} 259

\bibitem {60} Ritzinger P, Reichlova H, Kriegner D, Markou A, Schlitz R, Lammel M, Scheffler D, Park G H, Thomas A, \v{S}t\v{e}da P, Felser C, Goennenwein S T B and V\'{y}born\'{y} K 2021 {\it Phys. Rev. B} {\bf 104} 094406
\bibitem {61} Tu D, Wang C and Zhou J 2025 {\it Phys. Rev. B} {\bf 112} L041405
\bibitem {62} Dai Y, Zhao Y W, Ma L, Tang M, Qiu X P, Liu Y, Yuan Z and Zhou S M 2022 {\it Phys. Rev. Lett.} {\bf 128} 247202

\bibitem {63} Hodovanets H, Eckberg C J, Campbell D J, Eo Y, Zavalij P Y, Piccoli P, Metz T, Kim H, Higgins J S, Paglione J 2022 {\it Phys. Rev. B} {\bf 106} 235102
\bibitem {64} Gonzalez Betancourt R D, Zub\'{a}\v{c} J, Geishendorf K, Ritzinger P, R\r{u}\v{z}i\v{c}kov\'{a} B, Kotte T, {\v Z}elezn{\' y} J, Olejn{\' i}k K, Springholz G, B{\" u}chner B, Thomas A, V{\' y}born{\' y} K, Jungwirth T,Reichlov{\' a} H and Kriegner D 2024 {\it npj Spintronics} {\bf 2} 45
\bibitem {65} Lv S, Guo H, Qi Q, Li Y, Hu G, Zheng Q, Wang R, Si N, Zhu K, Zhao Z, Han Y, Yu W, Xian G, Huang L, Bao L, Lin X, Pan J, Du S, He J, Yang H and Gao H-J 2025 {\it Adv. Funct. Mater.} {\bf 35} 2412876

\bibitem {66} Cui Z, Grutter A J, Zhou H, Cao H, Dong Y, Gilbert D A, Wang J, Liu Y-S, Ma J, Hu Z, Guo J, Xia J, Kirby B J, Shafer P, Arenholz E, Chen H, Zhai X and Lu Y 2020 {\it Sci. Adv.} {\bf 6} eaay0114
\bibitem {67} Shi W, Zhang J, Chen X, Zhang Q, Zhan X, Li Z, Zheng J, Wang M, Han F, Zhang H, Gu L, Zhu T, Liu B, Chen Y, Hu F, Shen B, Chen Y and Sun J 2023 {\it Adv. Funct. Mater.} {\bf 33} 2300338


\bibitem {68} Stoner E C and Wohlfarth E P 1948 {\it Philos. Trans. R. Soc. Lond. Ser. A} {\bf 240} 599

\bibitem {69} {\v Z}elezn{\' y} J, Gao H, Manchon A, Freimuth F, Mokrousov Y, Zemen J, Ma{\v s}ek J, Sinova J and Jungwirth T 2017 {\it Phys. Rev. B} {\bf 95} 014403
\bibitem {70} Volny J, Wagenknecht D, {\v Z}elezn{\' y} J, Harcuba P, Duverger-Nedellec E, Colman R H, Kudrnovsk{\' y} J, Turek I, Uhl{\' i}{\v r}ov{\' a} K and V{\' y}born{\' y} K 2020 {\it Phys. Rev. Mater.} {\bf 4} 064403
\bibitem {71} Bozorth R M 1946 {\it Phys. Rev.} {\bf 70} 923

\bibitem {72} Perdew J P, Ruzsinszky A, Csonka G I, Vydrov O A, Scuseria G E, Constantin L A, Zhou X and Burke K 2009 {\it Phys. Rev. Lett.} {\bf 102} 039902
\bibitem {73} Bl\"{o}chl P E 1994 {\it Phys.Rev. B} {\bf 50} 17953
\bibitem {74} Kresse G and Furthm{\"u}ller J 1996 {\it Phys. Rev. B} {\bf 54} 11169
\bibitem {75} Dudarev S L, Botton G A, Savrasov S Y, Humphreys C J and Sutton A P 1998 {\it Phys. Rev. B} {\bf 57} 1505



\end{thebibliography}
\end{document}